\documentclass[journal=jacsat,manuscript=article]{achemso}

\usepackage[version=3]{mhchem} 

\author{Sabrina Juergensen}
\author{Moritz Kessens}
\affiliation{Department of Physics, Freie Universität Berlin, Berlin, Germany}
\author{Charlotte Berrezueta-Palacios}
\affiliation{Department of Physics, Freie Universität Berlin, Berlin, Germany}
\author{Nikolai Severin}
\author{Sumaya Ifland}
\author{Jürgen P. Rabe}
\affiliation{Department of Physics \& IRIS Adlershof, Humboldt-Universität zu Berlin, Berlin, Germany. }
\author{Niclas S. Mueller}
\affiliation{NanoPhotonics Centre, Cavendish Laboratory, Department of Physics, University of Cambridge, Cambridge, United Kingdom.}
\author{Stephanie Reich}
\email{stephanie.reich@physik.fu-berlin.de}
\affiliation{Department of Physics, Freie Universität Berlin, Berlin, Germany}

\title[An \textsf{achemso} demo]
  {Collective States in Molecular Monolayers on 2D Materials}

\abbreviations{IR,NMR,UV}

\begin{document}







\begin{abstract}
  Collective excited states form in organic two-dimensional layers through the Coulomb coupling of the molecular transition dipole moments. They manifest as characteristic strong and narrow peaks in the excitation and emission spectra that are shifted to lower energies compared to the monomer transition. We study experimentally and theoretically how robust the collective states are against homogeneous and inhomogeneous broadening as well as spatial disorder that occur in real molecular monolayers. Using a microscopic model for a two-dimensional dipole lattice in real space we calculate the properties of collective states and their extinction spectra. We find that the collective states persist even for 1-10\% random variation in the molecular position and in the transition frequency, with similar peak position and integrated intensity as for the perfectly ordered system. We measure the optical response of a monolayer of the perylene-derivative MePTCDI on two-dimensional materials. On the wide band-gap insulator hexagonal boron nitride it shows strong emission from the collective state with a line width that is dominated by the inhomogeneous broadening of the molecular state. When using the semimetal graphene as a substrate, however, the luminescence is completely quenched. By combining optical absorption, luminescence, and multi-wavelength Raman scattering we verify that the MePTCDI molecules form very similar collective monolayer states on hexagonal boron nitride and graphene substrates, but on graphene the line width is dominated by non-radiative excitation transfer from the molecules to the substrate. Our study highlights the transition from the localized molecular state of the monomer to a delocalized collective state in the two-dimensional molecular lattice that is entirely based on Coulomb coupling between optically active excitations of the electrons or molecular vibrations. The outstanding properties of organic monolayers make them promising candidates for components of soft-matter optoelectronic devices.
\end{abstract}

Highly ordered lattices of perylene derivatives or other flattened organic molecules on surfaces are intriguing systems to study molecular interactions, the interaction between molecules and substrates, light-matter coupling, and the emergence of collective molecular states.\cite{Gomez1998,Mueller2011,Zhao2019,Wuerthner2011,Alkhamisi2018}
The coupling between the transition dipoles of many non-covalently bound molecules leads to collective states with short life times, high emission rates, and narrow line widths.\cite{Bricks2017,Zhao2019,Mueller2022} Of particular interest are the optically active excitonic transitions in such two-dimensional (2D) molecular lattices that occur in the visible and near ultra-violet energy range. Their collective excitation is typically red-shifted compared to the monomer transition with a vanishing Stokes shift between the excitation and emission energy.\cite{Zhao2020, Zhao2019, Bricks2017} The bright light emission may be exploited in optoelectronic devices that will be tunable by molecular synthesis as well as by changes in the monolayer environment.\cite{Zhao2019,Zhao2020}

The Coulomb coupling of molecular transition dipoles is an interesting model to study the transition from localized molecular excitations to collective 2D states that are delocalized in space or the emergence of 2D Frenkel excitons, that were introduced in the framework of organic crystals.\cite{Davydov1971,Kasha1963,Czikkely1970,Yamagata2012,Zhao2019} Collective molecular excitations, however, also resemble the formation of collective plasmons in monolayers of metal nanoparticles and 2D arrays of atoms in optical lattices.\cite{Mueller2018,Vieira2019, Sherson2010,Masson2022,Schaefer2020} Molecular monolayers occupy a unique parameter space for such artificial 2D systems: The lattice constants of molecular monolayers are on the order of 1\,nm, which is much smaller than for plasmonic (10\,nm) and optical lattices (100-1000\,nm) allowing a translational periodicity that is two orders of magnitude smaller than the wavelength of light. The effects of energetic disorder and of the lattice environment are typically neglected in plasmonic and atomic 2D lattices, because the structures are either very precisely controlled and isolated or are large enough to be less sensitive to imperfections.\cite{Schulz2020,Barredo2016} This situation is very different for molecular lattices. Molecular transitions show strong inhomogeneous broadening in ambient environments ($>$\,10\% of their transition frequency) resulting in lattices that are composed of different transition dipoles. Molecular self organization during growth is mainly driven by weak intermolecular forces and by the interaction with the substrate making molecular 2D lattices potentially more prone to disorder than plasmonic structures and optical lattices.\cite{Mueller2013,Dienel2008} The presence of the substrate screens dipole coupling and may quench excited molecular states providing a strong non-radiative decay channel.\cite{Kim2009,Matte2011} The question arises how robust collective states are against spatial and energetic disorder and how quenching affects its formation, energetic position, and line width. 

Here, we study collective molecular states in organic 2D lattices of perylene-derivatives on different 2D materials. Using a microscopic theory of collective dipoles, we explain our experimental observations and examine the dependence of the collective state on homogeneous and inhomogeneous broadening. Experimentally we realize 2D molecular lattices by growing N,N'-Dimethyl-3,4,9,10-Perylentetracar\-boxylicdiimide (MePTCDI) on multi-layer hexagonal boron nitride (hBN) and few-layer graphene, where it forms a 2D square lattice. On hBN, the collective MePTCDI state shows a red shift of 60\,meV and 60\% reduction in the full width at half maximum (FWHM) compared to the monomer transitions, which agrees with the microscopic description. We show that the collective state is also present on graphene as a conductive material although light emission is quenched by five orders of magnitude, but it manifests in optical absorption and as a resonance in the Raman response. Using a microscopic model of collective dipoles we show how the energetic position of the collective eigenstate depends on the number of interacting molecules, their properties, and packing density. The state is robust against molecular quenching and variations in  position and transition frequencies. These factors lead to a broadening of the collective state but little or no shift in its peak position.

\section{Results and Discussion}
\subsection{Theory of Collective Molecular Excitations}

\begin{figure*}
    \includegraphics[width=11cm]{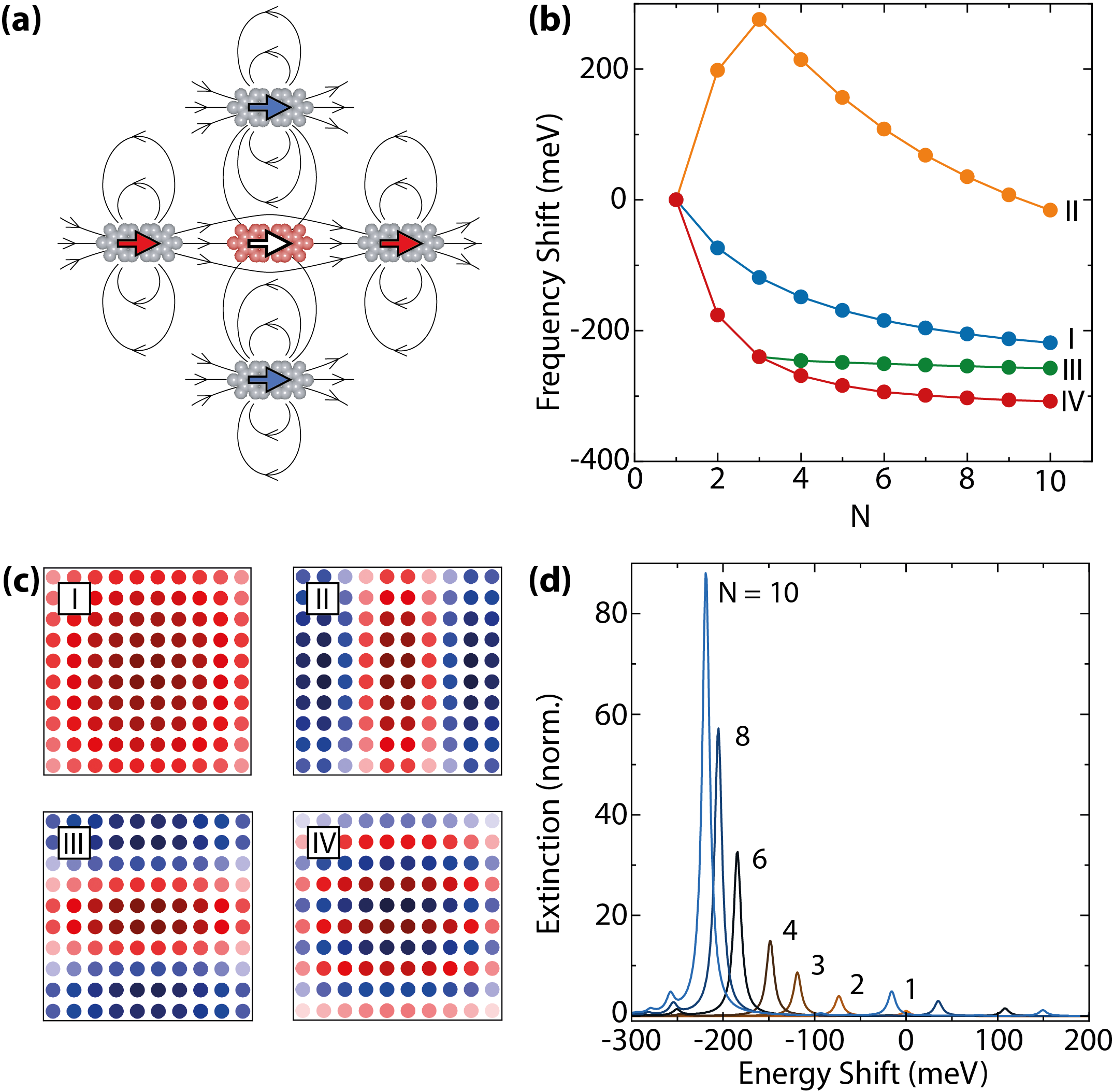}
    \caption{Microscopic model of collective excitons. \textbf{(a)} Nearest-neighbor interactions of dipoles (bold arrows) through their electric fields (black lines). Two neighbors (red arrows) are in an attractive interaction with the centered dipole (white arrow), whereas the other two neighbors are in an anti-bonding configuration (blue arrows). \textbf{(b)} Energy shift of selected eigenstates with eigenvectors shown in (c) for an $N\times N$ square lattice. Eigenstate I (blue) has the strongest dipole moment, eigenstate IV (red) is optically inactive, but has the largest energy shift. The labels connect the eigenenergies to the eigenvectors in (c). \textbf{(c)} Selected collective eigenmodes of a molecular $10\times10$ lattice in real space. Red (blue) dots indicate positive (negative) dipole moment along $x$. \textbf{(d)} Collective extinction of an $N \times N$ square lattice of dipoles with increasing $N$ (see labels). The spectra are normalized to the extinction of a single dipole with $\omega_0=2\,$eV, $d_0=10\,\mathrm{D}$ and $\gamma_0 = 5\,\mathrm{meV}$. Lattice constant $a = 1\,\mathrm{nm}$, $\epsilon_m = 1$, and dipole polarization along $x$ in all simulations.}
    \label{fig:CollectiveDipoleTheory}
\end{figure*}

Collective states in molecular dimers and small aggregates are a well-known phenomenon for molecules in solution, where they arise from stacking and alignment of molecular transition dipoles.\cite{Kasha1961,Czikkely1970,Bricks2017,Hestand2018} A molecular dimer with a head-to-tail configuration of its transition moments has a red-shifted transition energy ($J$-aggregate) and shows  the characteristic optical properties of a superradiant state that has a larger transition  moment than the monomer.\cite{Zhao2019,Deshmukh2022} The side-by-side configuration, in contrast, results in a blue-shift of the optically active state ($H$-aggregates), and increases the difference between absorption and emission spectra (Stokes shift).\cite{Zhou2018} The concept of pure $J$ and $H$ configuration is easily extended from a dimer to one-dimensional molecular chains, but not to 2D lattices that necessarily combine head-to-tail and side-by-side arrangements, see Fig.~\ref{fig:CollectiveDipoleTheory}a for a sketch considering the nearest neighbors of a molecule. Sometimes the 2D configuration is called an $HJ$-aggregate or is extended to the intermediate $I$-aggregate that includes other types of interactions in addition to dipole-dipole coupling.\cite{Yamagata2012, Hestand2018, Deshmukh2022, Deshmukh2019,Kirstein1995}

To model collective molecular states in 2D systems we consider a finite 2D arrangement of molecules that interact through their transition dipole moments.\cite{Orrit1986,Fidder1991,Bettles2017} We use a real space model for its ability to describe any arrangement of dipoles in space without imposing, \textit{e.g.}, translational symmetry. Each molecular transition is represented by a point dipole $d=\alpha E_0$ induced by the field $E_0$ with the polarizability
\begin{equation}
    \alpha = \frac{d_{ge}^2}{\hbar(\omega_0 - \omega  - \mathrm{i}\gamma_0)},
\end{equation}
where $\omega$ is the driving frequency, $\omega_0$ the molecular transition frequency, $2 \gamma_0$ its spectral broadening, and $d_{ge}$ the transition dipole moment. Each individual dipole interacts \textit{via} its electric field with all other transition dipoles. The interaction changes the dipole moment at each lattice site and gives rise to collective dipolar eigenmodes $\textbf m_p$ with polarizability $\alpha_p$, see Methods for details. The individual polarizability $\alpha$ is replaced by the collective lattice polarizability \cite{Bettles2017}
\begin{equation}
    \alpha_\mathrm{coll}=\sum_p\alpha_p=\sum_p \frac{d_{ge}^2}{\hbar(\omega_0-\omega+\Delta_p)-i\hbar(\gamma_0+\gamma_p)}.
\end{equation}
The collective transitions are shifted in frequency by $\Delta_p=d_\mathrm{ge}^2\mathrm{Re}(g_p)$ compared to $\omega_0$, where $g_p$ is the complex eigenvalue of the Green's tensor describing the near- and far-field dipole-dipole coupling, see Methods. In addition, the decay constant is increased by $\gamma_p=d_{ge}^2\mathrm{Im}(g_p)$, which is the characteristic increase in the emission rate, observed in molecular aggregates. The collective response of the molecular lattice can be measured experimentally by the extinction and compared to a calculated extinction coefficient that depends on the sum over $\alpha_p$, see Methods for details. Supplementary Figure\,1 shows dipole chains (1D) of increasing length as calculated with the microscopic dipole model reproducing the well-known frequency shifts of pure $J$- and $H$-aggregates.

In Fig.~\ref{fig:CollectiveDipoleTheory}b we show the shift in the energy of selected collective states in an $N\times N$ square lattice. The absolute magnitude of the shift depends on the collective eigenvector, the individual transition dipoles, the packing density, lattice size and type, and dielectric screening. The eigenenergies shift with $N$ but for most modes saturate for 100\,dipoles in the finite lattice ($N=10$), Fig.~\ref{fig:CollectiveDipoleTheory}b. The eigenmode with the strongest dipole moment, eigenvector I in Fig.~\ref{fig:CollectiveDipoleTheory}c, has all dipoles oriented parallel resulting in a maximum shift of 200\,meV or 10\% of the transition frequency. The mode with the second strongest dipole moment corresponds to an anti-bonding configuration with three stripes of dipoles that are aligned side-by-side (vector II). For very small lattices it results in a strong blue shift of the collective state, because of the purely repulsive nearest-neighbour interactions for $N=2$ and $3$. With increasing lattice size ($N\ge 4$), the dipole-dipole coupling along the $x$ direction becomes partly attractive (parallel dipoles along $x$) and partly repulsive (antiparallel  along $x$ and parallel along $y$). The frequency shift of eigenvector II  converges towards zero for larger $N$ since all attractive and repulsive contributions cancel approximately. Interestingly, eigenvector III with a corresponding pattern varying along the $y$ axis has a red shift of $\sim 300$\,meV, Fig.~\ref{fig:CollectiveDipoleTheory}b, \textit{i.e.}, higher than for the perfectly parallel dipoles in eigenvector I. When all dipoles oscillate in-phase along $x$ the coupling is attractive for dipoles along the oscillation direction but repulsive perpendicular to it, Fig.~\ref{fig:CollectiveDipoleTheory}a. Modes with anti-parallel stripes of dipoles perpendicular to the oscillation direction, therefore, have a stronger binding contribution and a larger frequency shift. The mode with the highest energy shift, eigenvector IV, has single lines with alternating polarization direction resulting in the configuration with the strongest bonding character along the lines and perpendicular to them. This means that for this 2D square lattice the excited state with the lowest energy is dipole forbidden and luminescence needs to be activated thermally.\cite{Yamagata2012}

Absorption and emission of the perfect 2D dipole lattice is dominated by a few bright modes, while the majority of the eigenvectors are dipole forbidden.\cite{Wuerthner2011} Eigenvector I has a dipole moment of 94\,D, which increases its radiative decay $\gamma_p$ by a factor of 88 compared to the individual dipole ($N=10$). For mode II and III the increase amounts to a factor of five and two, respectively. The increase in the radiative decay by approximately $\Gamma_\mathrm{rad}(\omega) \propto N^2$ is known as the Dicke superradiance \cite{Dicke1954} that arises from the collective emission of many molecules. While the rate of spontaneous emission increases for specific modes in Fig.~\ref{fig:CollectiveDipoleTheory}c, the overall integrated intensity remains the same for arrays of coupled and uncoupled dipoles. The absorption or extinction spectrum of the 2D lattice is governed by the total response of all collective eigenstates resulting in a single dominant peak at the energy of the collective mode with the strongest dipole moment, Fig.~\ref{fig:CollectiveDipoleTheory}d. The integrated intensity of the extinction peak follows $\sigma_\mathrm{ext} \propto 0.9 N^2$. This means that the extinction spectrum of a 2D lattice is of similar intensity for coupled and uncoupled transition dipoles, because individual\mbox \:dipole intensities have an overall extinction $\sigma_\mathrm{ext} \propto N^2$ through the sum of all uncoupled contributions. Nevertheless, the peak position is a clear fingerprint for a collective state.

\begin{figure*}
    \includegraphics[width=16cm]{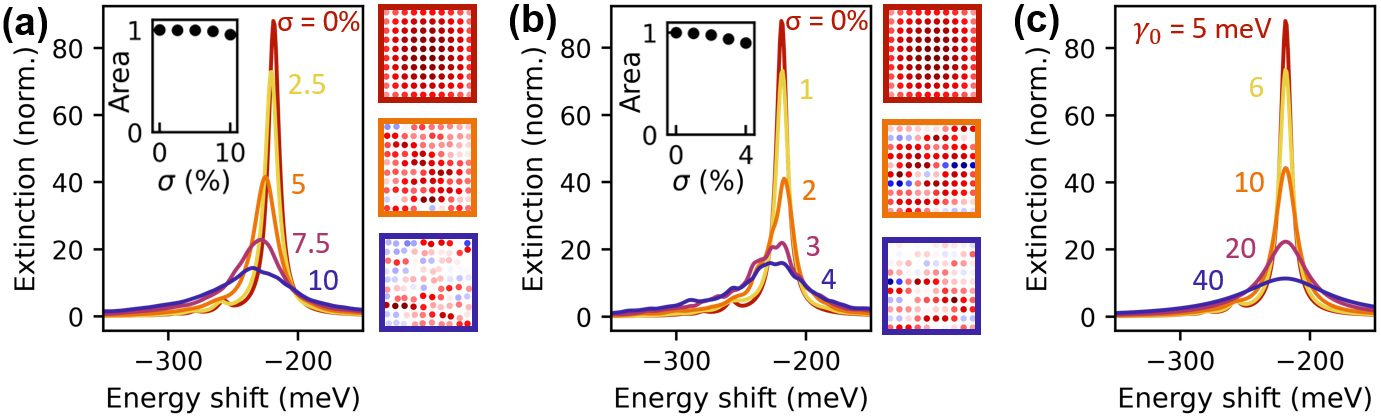}
    \caption{Disorder in 2D dipole lattices. \textbf{(a)} Effect of position disorder on collective extinction. Average spectra are calculated from $1000$ random lattices with standard deviation $\sigma$ (units of lattice constant) of each dipole position in the $xy$ plane. The panels to the right show dominant collective modes of selected random lattices with $\sigma = 0\%$, $5\%$, and $10\%$ from top to bottom. \textbf{(b)} Effect of frequency disorder on collective extinction of $10\times 10$ square lattice. The panels to the right show dominant collective modes for selected random lattices with $\sigma = 0\%$, $2\%$, and $4\%$ from top to bottom. Insets in (a) and (b) give integrated peak areas. \textbf{(c)} Collective extinction for perfect lattice with increasing homogeneous broadening $\gamma_0$. All spectra are referenced to the extinction of a single dipole with $d_0=10\,\mathrm{D}$ and $\gamma_0 = 5\,\mathrm{meV}$. Further parameters: lattice constant $a = 1\,\mathrm{nm}$, $\epsilon_m = 1$, and dipoles along $x$.}
    \label{fig:Disorder}
\end{figure*}

The formation of the collective 2D state, its energetic position, and even the intensity of the extinction spectrum are surprisingly robust against spatial disorder and inhomogeneous broadening of the molecular transition as we show now with our real space simulations, Fig.~\ref{fig:Disorder}.\cite{Fidder1991} We first consider spatial disorder that may arise from variations of the dipole positions so that the dipole monolayer deviates from the perfect 2D lattice. We model this by varying the dipole positions $\mathbf{r}_i + \delta\mathbf{r}_i$ in the $xy$ plane around the sites $\mathbf{r}_i$ of the perfect lattice, where $\delta\mathbf{r}_i$ is obtained through random sampling from a multivariate Gaussian distribution. To account for the large sample area that is typically probed in experiments we averaged the extinction spectrum over several random lattices, Fig.~\ref{fig:Disorder}a. 
Disorder increases the width and decreases the maximum intensity of the extinction spectrum, but the peak area remains within 95\% of the original intensity for a standard deviation of $\sigma=10\%$, inset in Fig.~\ref{fig:Disorder}a. Collective states continue to form despite the spatial disorder, but the delocalized eigenvectors of the perfect lattice become more localized with disorder, right panels in Fig.~\ref{fig:Disorder}a. The frequency shift of the collective excitation increases slightly in the disordered lattice, Fig.~\ref{fig:Disorder}a, because the dipole-dipole interaction scales with $1/\vert \mathbf{r}_i - \mathbf{r}_j \vert^3$.\cite{Ravets2014} A (random) decrease in the dipole distance causes a larger red shift than the blue shift induced by the corresponding increase in dipole distance.

Another source of disorder is inhomogeneous broadening or fluctuations in the transition frequency from one dipole to the next. For most dye molecules the luminescence and absorption linewidths under ambient conditions at room temperature ($\approx 10 - 100$\,meV) are dominated by inhomogeneous broadening compared to the much smaller radiative decay constants ($\hbar\gamma_\mathrm{rad} \approx 10^{-8}-10^{-6}$\,eV). At first sight, this appears to be a more serious distortion for the formation of a collective state that depends on the Coulomb coupling of transition dipoles (or the absorption and emission of virtual photons). However, the collective state remains present despite random fluctuations in excitation frequencies, see Fig.~\ref{fig:Disorder}b and Methods for details on the simulations. While the dominant collective mode continues to be delocalized in the lattice, it is formed by a subset of dipoles, right panels in Fig.~\ref{fig:Disorder}b. As for spatial disorder, inhomogeneous broadening has little effects on the integrated intensity (90\% intensity for 4\% disorder). Although the width of the collective extinction peak increases with the inhomogeneously broadened dipoles, its line width is smaller than expected from the variations in the individual peak positions, an effect that is known as motional narrowing in molecular aggregates.\cite{Knapp1984} We fit the spectra in Fig.~\ref{fig:Disorder}b with a single Lorentzian and determined the inhomogeneous contribution to the FWHM. The inhomogeneous contribution in the $10\times10$ dipole lattice was only 30\% of the inhomogeneous broadening for the individual dipoles, because the collective eigenvectors combine dipoles of similar frequency in the formation of the collective state, see right panels in Fig.~\ref{fig:Disorder}b. 

Finally, we show the effect of an additional non-radiative decay channel or homogeneous broadening by varying $\gamma_0$ in Fig.~\ref{fig:Disorder}c. This additional contribution affects all individual dipoles in exactly the same way and leads to an incerease in the FWHM of the collective eigenmodes. The cooperative frequency shift $\Delta_p = d_\mathrm{ge}^2 \mathrm{Re}(g_p)$, on the other hand, depends only on the transition dipole $d_\mathrm{ge}$ and the geometry of the lattice \textit{via} $\mathrm{Re}(g_p)$. It is independent of $\gamma_0$ as confirmed by the constant peak position in Fig.~\ref{fig:Disorder}c. 

To summarize, we simulated a molecular monolayer by a square lattice of interacting transition dipoles in real space. The interaction between the molecules leads to collective eigenstates that give rise to a strong red shift of the predicted excitation frequencies. This process is robust against spatial and energetic disorder, because of the strong dipole-dipole coupling in the tightly packed molecular layers. We also find that an increase in non-radiative decay will not affect the collective frequency that for a perfect lattice depends only on the individual transition frequency, the strength of the transition dipole, the 2D lattice type, its lattice constant, and the screening by the environment. We now realize such 2D molecular lattices experimentally to study the collective states by optical spectroscopy.

\subsection{Growth and Structure of MePTCDI Monolayers}

\begin{figure}
\centering
\includegraphics[width=0.6\textwidth]{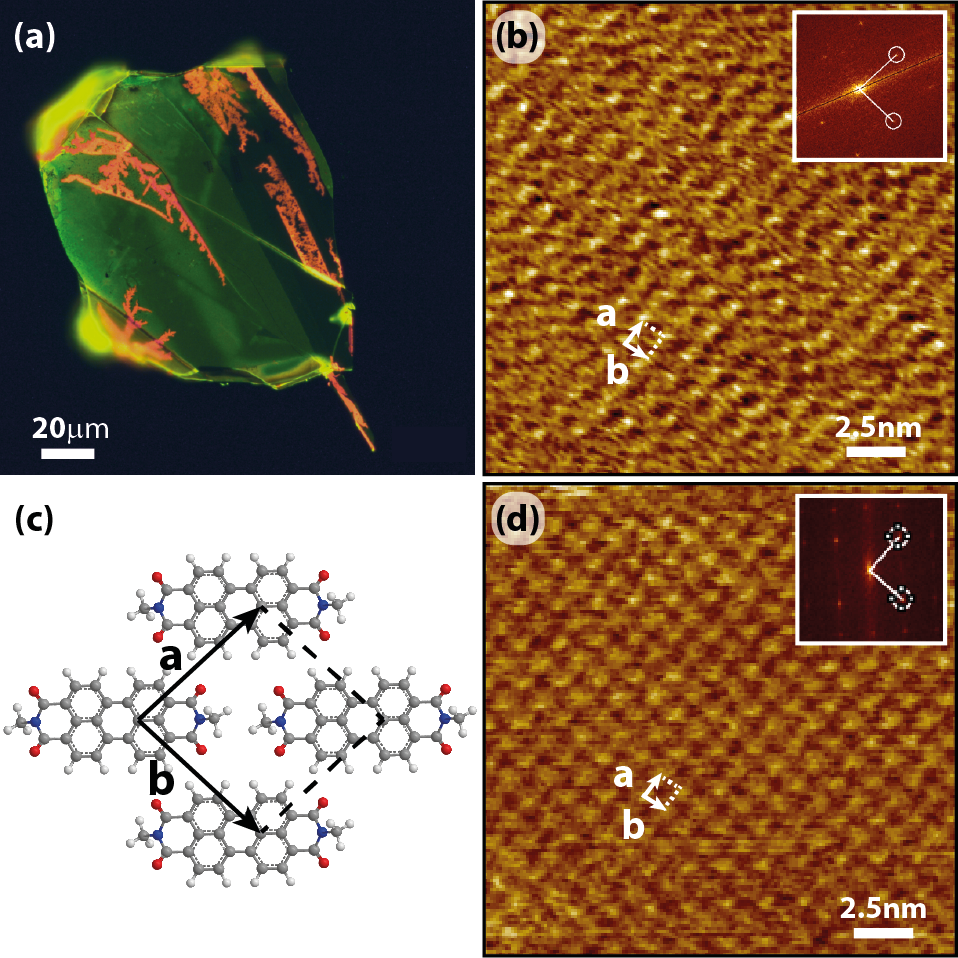}
\caption{\label{fig:AFM} Structure of the MePTCDI monolayer on multi-layers of hBN and few-layer graphene. \textbf{(a)} Fluorescence microscopy image of an hBN flake with ordered MePTCDI molecules on top. The area with green luminescence belongs to highly ordered monolayers of MePTCDI, whereas red luminescence are molecular aggregates (bulk). High resolution AFM phase images of the grown molecular layer on multi-layer \textbf{(b)} hBN and \textbf{(d)} graphene. Both images show a periodic nearly square pattern. The insets are FFT patterns of the phase images in (b) and (d), confirming the periodicity of the monolayers. \textbf{(c)} Brick stone lattice structure of the organic MePTCDI layer with the unit cell shown by black lines ($a=b=11.8$~\AA, $\angle{\mathbf{a},\mathbf{b}}=(84\pm2)^\circ$) agreeing within the experimental uncertainty of the unit cell from AFM imaging. The grey spheres of the molecule structure are carbon, red oxygen, blue nitrogen, and white hydrogen atoms. }
\end{figure}

Monolayers of MePTCDI were grown on multi-layers of hBN as an insulating and few-layer graphene as a conductive substrate, see Methods. MePTCDI is a planar dye molecule that belongs to the perylene family with a conjugated $\pi$-system, see molecular structure in Fig.~\ref{fig:AFM}c. It was previously shown to form micron-sized monolayers on atomically flat hBN.\cite{Zhao2019} We initially characterize the MePTCDI structure on hBN with fluorescence microscopy and high-eigenmode tapping mode atomic force microscopy (AFM), see Fig.~\ref{fig:AFM}a and Supplementary Fig.~2.\cite{Severin2022} Figure~\ref{fig:AFM}a shows a multi-layer hBN flake with MePTCDI molecules on top that were deposited by physical vapor deposition, see Methods. The green luminescence is emitted from an MePTCDI monolayer that arranged non-covalently on the hBN substrate.\cite{Zhao2019} It has an emission maximum at 2.25\,eV (551\,nm) as we will discuss in detail in the next Section. The monolayer areas dominate the sample in Fig.~\ref{fig:AFM}a and account for 84\% of the hBN area. The red areas (16\%) are molecular aggregates where the molecules stacked in 3D structures and interact \textit{via} $\pi-\pi$ coupling in addition to the Coulomb interaction.\cite{Zhao2019} The agglomerate fluorescence spectrum peaks at 1.75\,eV (709\,nm), Supplementary Fig.~3, resulting in the red appearance in the fluorescence microscope image. 

We determined the structure of the MePTCDI monolayers using high-resolution AFM, see Fig.~\ref{fig:AFM}b and Supplementary Fig.~4 for a large scale image at lower resolution. The image shows an almost square lattice with lattice constants $a=b=(11.8\pm0.3$)\,\AA \:and an angle $\angle{\mathbf{a},\mathbf{b}}=(84\pm2)^\circ$ as determined from a fast Fourier transform (FFT) of the AFM image, see inset. Figure~\ref{fig:AFM}c sketches the obtained monolayer structure that corresponds to the so-called brick stone lattice.\cite{Saikin2013} It results from the orientation of the transient molecular dipole moments plus the repulsion of the positively charged oxygen atoms. Dipole-dipole coupling aligns the molecules in a line along their long axis. The next row of molecules is placed in parallel but shifted for maximum distance between the oxygens on neighboring MePTCDI molecules. Multi-layers of MePTCDI form a herringbone structure,\cite{Zhao2019} but the brick stone lattice in Fig.~\ref{fig:AFM}b,c has denser packing and is, therefore, favored in the monolayer. The MePTCDI monolayer structure on multi-layer graphene, Fig.~\ref{fig:AFM}d, is identical within the experimental error to the one on hBN, Fig.~\ref{fig:AFM}b. The substrate had no direct effect on the in-plane structure of the molecular monolayer, which is reasonable because of the similar interaction and coupling strength of aromatic molecules to graphene and hBN.\cite{Berland2013} The structure of the MePTCDI monolayer is determined by intermolecular interactions \textit{via} Coulomb coupling and oxygen repulsion; the two-dimensional crystals only ensure the flat arrangement. Due to their structural similarity, the two MePTCDI monolayers are excellent candidates to study how the interaction with the substrates affects the molecular transitions and the formation of collective states.

\subsection{Collective MePTCDI Exciton}
\label{CollExcitation}

\begin{figure}
\centering
\includegraphics[width=0.65\textwidth]{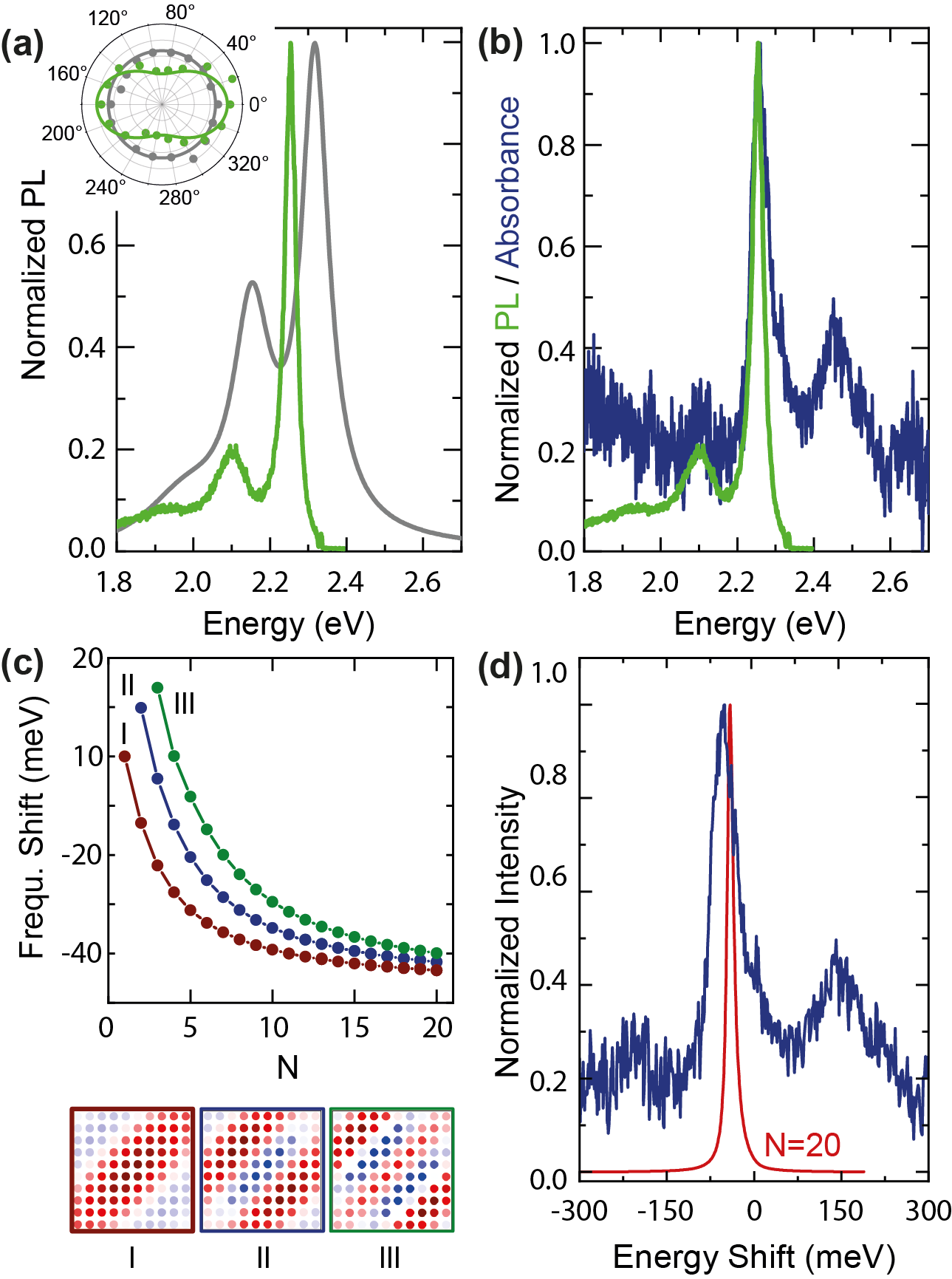}
\caption{\label{fig:PL_ABS} Fluorescence and absorption of the MePTCDI monolayer. \textbf{(a)} Fluorescence spectra of an MePTCDI monolayer (green) and MePTCDI in dimethyl sulfoxide (grey). For better comparison the emission and absorption spectra were normalized to one. The inset shows the polarization dependence of the light emission using the same color coding. \textbf{(b)} Comparison of the micro-fluorescence (green) and micro-absorption (blue) spectrum of the MePTCDI monolayer. Showing only a very small Stokes shift. \textbf{(c)} Frequency shift of collective eigenstates of an $N\times N$ brick stone lattice on hBN. Eigenstate I (red) has the largest net dipole, eigenstate II (blue) the second largest and eigenstate III (green) the third largest net dipole. The panels below show the collective modes of the eigenstates in (c) for $N = 10$. Red (blue) dots indicate positive (negative) dipole moment along the diagonal. \textbf{(d)} Comparison of the micro-absorption (blue) spectrum  and the simulated spectrum (red) of a $20\times20$ brick stone lattice.}
\end{figure}

Light absorption and emission from the MePTCDI monolayer at room temperature, Fig.~\ref{fig:PL_ABS}, has a strong peak at 2.25\,eV, which originates from collective molecular states. It is shifted by 60\,meV to smaller energies compared to the monomer (2.31\,eV), Fig.~\ref{fig:PL_ABS}a. This is in good agreement with Zhao \textit{et al.}.\cite{Zhao2019} The emission is polarized, inset in Fig.~\ref{fig:PL_ABS}a, which confirms optically that the molecules form a highly orientated lattice. In addition to the red shift of the emission, we find a vanishing Stokes shift in the monolayer absorption and emission, Fig.~\ref{fig:PL_ABS}b. The Stokes shift of MePTCDI is already quite small in solution (60\,meV), but the lower flanks of the absorption and emission peak are identical in the monolayer.\cite{Engel2006} Other signatures of the collective state are the increase in the dominant zero-phonon-line compared to the phonon sideband and the much narrower line width of the monolayer emission compared to the molecules in solution, Table~\ref{tab:energie}.\cite{Spano2010,Bricks2017}

We calculated the expected frequency shift for a 2D lattice of MePTCDI monomers with $\hbar\omega_0=2.31\,$eV, $\hbar\gamma_0=5\,$meV, $d_{ge}=8.8$\,D, and $\epsilon_m=2.7$, Fig.~\ref{fig:PL_ABS}c,d. The dipoles were oriented at $45^\circ$ in a square lattice as dictated by the brick stone structure of our experiments, Fig.~\ref{fig:AFM}c. In this lattice the eigenmode with the largest frequency shift (eigenvector I) has also the largest dipole moment, Fig.~\ref{fig:PL_ABS}c. The luminescence by the MePTCDI lattice is therefore not thermally activated and remains present at low temperature which is in good agreement with prior experiments.\cite{Zhao2019} As the calculated frequency shift we obtain a red shift of 40\,meV for the collective MePTCDI state, which is smaller than the experimental shift (60\,meV), Fig.~\ref{fig:PL_ABS}c,d. However, the simulation has a number of uncertainties: First, the molecules are large compared to their distance, which means that the point dipole approximation is not strictly valid for this configuration. Second, $\hbar\omega_0$ was measured in solution and might actually differ on a solid substrate. Also, we assumed the screening by the hBN substrate to yield an effective background dielectric constant of $n=1.64$, which corresponds to half space filling by the substrate and may overestimate the screening within the layer, see Supplementary Information. 

The emission spectrum of the MePTCDI monolayer has a narrow line width (FWHM = 36\,meV) with a slightly larger width (45\,meV) in absorption. The absorption is broader because all states with finite dipole moment contribute to excitation, but light emission occurs predominantly from the lowest-lying optically active states. Despite its narrow appearance, the line width of the collective MePTCDI state remains dominated by inhomogeneous broadening with little contribution from spatial disorder. The FWHM of the monolayer amounts to 40\% of the molecular transition in solution, Table~\ref{tab:energie}, in reasonable agreement with the predicted narrowing (30\%, Fig.~\ref{fig:Disorder}b). The dominance of inhomogeneous broadening is also confirmed by time-resolved measurements that reported a lifetime of 30\,ps and nearly 100\% quantum yield for the monolayer,\cite{Zhao2019} which yields a lifetime limited FWHM $\approx 30\,\mu$eV, \textit{i.e.}, three orders of magnitude below the observed width. 

\begin{table}[h]
    \centering
    \caption{Transition energies and linewidth of the individual and collective MePTCDI excitons.}
    \vspace{0.4cm}
    \begin{tabular}{lcc}\hline
sample & $\omega$ (eV)  &  FWHM (meV)\\\hline
MePTCDI monomer &   2.31 &  90 \\
MePTCDI/hBN absorption &  2.25  & 45\\
MePTCDI/hBN emission   &   2.25 & 36\\
MePTCDI/graphene absorption & 2.26  & 96\\
MePTCDI/graphene Raman res. &  2.26  & 140\\\hline
    \end{tabular}
    \label{tab:energie}
\end{table}

\begin{figure}
\centering
\includegraphics[width=0.99\textwidth]{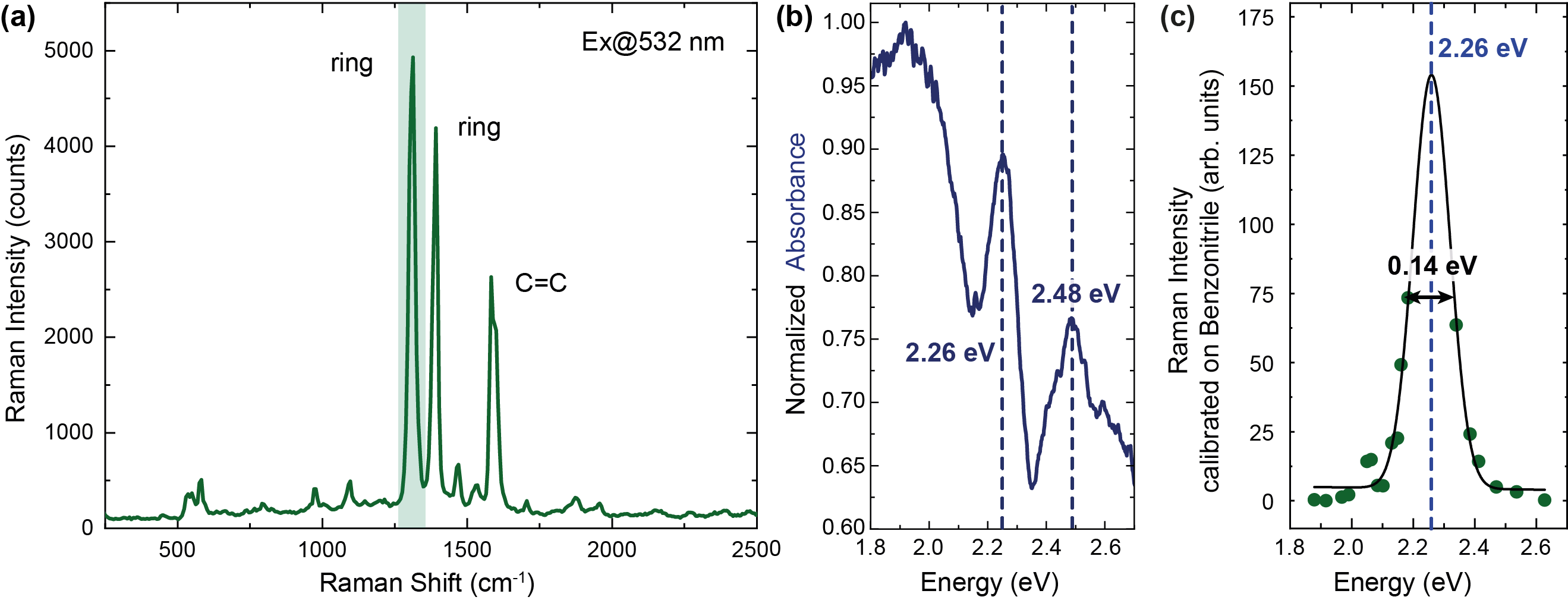}
\caption{\label{fig:RAMAN} MePTCDI monolayer on graphene. (a) Raman spectrum of the MePTCDI monolayer on graphene at 532\,nm excitation wavelength. The Raman modes of the dye molecule are indicated. (b) Normalized micro-absorbance spectrum of the MePTCDI monolayer on graphene. The spectrum has two absorption peaks that are located at 2.26\,eV and 2.48\,eV. (c) Resonant Raman profile of the ring stretch mode at 1309\,cm$^{-1}$. The molecular resonance is located at 2.26\,eV (549\,nm) and has a FWHM of 0.14\,eV.  }
\end{figure}

On graphene the characteristic MePTCDI monolayer emission vanishes. Instead we observe the Raman spectrum of the monolayer, Fig.~\ref{fig:RAMAN}a. The peaks at 1309\,cm$^{-1}$ and 1392\,cm$^{-1}$ correspond to ring stretch modes of the perylene core while the mode at 1588\,cm$^{-1}$ belongs to C=C stretching of the carbon rings.\cite{Akers1988} We attribute the quenching of the luminescence to a Förster resonant energy transfer that is very efficient, because of the small distance ($0.3\,$nm) between a planar dye and graphene, the parallel alignment of the transition dipoles, and the broadband absorption of multi-layer graphene in the visible and near IR. Experimentally, the intensity loss is at least five orders of magnitude or $\gamma_0/\gamma_{m\rightarrow G}\approx10^{-5}$, where $\gamma_{m\rightarrow G}$ is the rate of excitation transfer from the molecule into graphene and $\gamma_0$ the intrinsic molecular decay rate. 
 
To verify that the collective state of MePTCDI exists in the presence of graphene as a strong quenching agent, we measure the excitonic transition by absorption and resonant Raman scattering, Fig.~\ref{fig:RAMAN}b,c. The MePTCDI absorption on graphene has a main peak at essentially the same energy as on hBN but twice its line width, see Table~\ref{tab:energie} and Fig.~\ref{fig:RAMAN}b. The small shift in the collective frequency between hBN and graphene is explained by the difference in the screening strength. The increase in FWHM implies an additional broadening by a non-radiative channel with $\gamma_{nr}\approx 42\,$meV. Assuming $\gamma_{nr} = \gamma_{m \rightarrow G}$ and a lifetime of the MePTCDI transition on the order of 1\,ns (Refs.~\cite{Zhao2019,Lunedei2003}) we estimate a relative transfer $\gamma_0/\gamma_{m\rightarrow G}\approx 10^{-5}$, which agrees with the intensity loss of the MePTCDI monolayer on graphene.

In resonant Raman scattering we determined the integrated intensity of the ring stretch mode at 1309\,cm$^{-1}$ as a function of excitation energy, Fig.~\ref{fig:RAMAN}c. The profile has a maximum at 2.26\,eV with a FWHM of 140\,meV that corresponds to the incoming resonance with the collective MePTCDI state. We did not observe an outgoing resonance in the profile that is expected at the incoming resonance plus the phonon frequency.\cite{Yu2010,Wasserroth2018} Nevertheless, the Raman resonance clearly demonstrates that the collective dipole state exists for MePTCDI on graphene, although its emission is suppressed. Independent of the dielectric environment and excitation transfer, the molecular monolayer forms a collective state on flat surfaces. This remains true although in our system the broadening exceeds the coupling-induced red shift of the molecular excitonic transition of the MePTCDI monolayer.

\section{Conclusion}

In conclusion we studied the transition of a localized excitation in a molecular monomer to a delocalized collective state in a molecular monolayer considering homogeneous and inhomogeneous broadening and spatial disorder. We presented a model based on the point-dipole approximation to calculate the eigenstates of 2D lattices and their extinction spectra. The interaction between the transition dipoles leads to collective states in the lattice and a red-shifted extinction spectrum. The shift depends on the radiative lifetime of the monomer (or the transition dipole) and the packing density of the molecular lattice. We found that spatial and energetic disorder as well as homogenenous non-radiative broadening lead to an increase in the linewidth of the collective state but hardly affect its transition frequency. Due to the strong dipole interaction and the small distance of molecules in a typical 2D lattice, the collective state is very robust. To study the predicted behavior experimentally, we grew monolayers of MePTCDI on multi-layer hBN and few-layer graphene substrates. We observed a brick stone lattice in the monolayer on 2D materials with a single molecule in the 2D unit cell. The collective monolayer state was shifted by 60\,meV compared to the monomer transition which occurred on both hBN and graphene. On hBN we observed strong luminescence with characteristic signatures of superradiance like a narrowing of the transition and a vanishing Stokes shift. On graphene, the energy transfer from the MePTCDI molecules to the substrate led to an additional homogeneous broadening of 85\,meV, but the collective state remained at the same energy. Our study shows that collective states in 2D molecular lattices are robust against disorder that result in additional homogeneous and inhomogeneous broadening. This paves the way for superradiant devices using soft materials with scalable fabrication techniques and solution-based processing.

\section{Methods}

\subsection*{Microscopic Model of Collective Dipoles}
\label{MicroscopicModel}

We describe the excitons of each molecule as point dipoles $d = \alpha E_0$ that are excited by an external electric field $E_0$. 
\begin{equation}
    \alpha = \frac{d_{ge}^2}{\hbar(\omega_0 - \omega  - \mathrm{i} \gamma_0)}
\end{equation}
is the polarizability of each individual dipole, with $\omega$ the driving frequency, $\omega_0$ the exciton frequency, $2 \gamma_0$ its spectral broadening, and $d_{ge}$ the transition dipole moment. 
In a 2D lattice, the individual dipole moments are modified by the coupling to the electric fields of other dipoles (Fig.\ \ref{fig:CollectiveDipoleTheory}a), which changes the dipole moment at lattice site $i$ to
\begin{equation}
\label{eq:CoupledDipoleEquation}
    d_i = \alpha E_0(\mathbf{r}_i) + \alpha \sum_{i \neq j} \mathrm{G}_{ij} d_j,
\end{equation}
where $\mathrm{G}_{ij} \equiv \mathrm{G}(\mathbf{r}_{ij})$ is a Green function that accounts for near-field and far-field coupling, with 
\begin{equation}
\begin{split}
    \mathrm{G}(\mathbf{r}_{ij})\mathbf{d}_j =& \frac{k^3}{4\pi\epsilon_0\epsilon_m}\mathrm{e}^{\mathrm{i}k r_{ij}} \Bigg[\left( \frac{1}{k r_{ij}} + \frac{\mathrm{i}}{(k r_{ij})^2} - \frac{1}{(k r_{ij})^3} \right)\mathbf{d}_j  
    \\& - \left( \frac{1}{k r_{ij}} + \frac{3\mathrm{i}}{(k r_{ij})^2} - \frac{3}{(k r_{ij})^3} \right) (\hat{\mathbf{r}}_{ij}\cdot \mathbf{d}_j)\hat{\mathbf{r}}_{ij} \Bigg].
\end{split}
\end{equation}
Here $\mathbf{r}_{ij} = \mathbf{r}_i - \mathbf{r}_j$, $r_{ij}=\vert \mathbf{r}_{ij} \vert$, and $\epsilon_m$ is the dielectric screening by the substrate.

We assume that all dipoles oscillate along the same axis and therefore omit vector notation. Following Ref.~\cite{Bettles2017} we write Eq.\ (\ref{eq:CoupledDipoleEquation}) as $\mathbf{E}_0 = \mathrm{M} \mathbf{d}$, with a coupling matrix $\mathrm{M}$ that has the general form $\mathrm{M}_{ij} = \delta_{ij} \alpha^{-1} - (1-\delta_{ij})\mathrm{G}_{ij}$. The entries of the vectors $\mathbf{E}_0$ and $\bf d$ stand for each lattice site. To understand the collective behavior, we use an eigenmode expansion
\begin{equation}
    \mathrm{M} \mathbf{m}_p = \mu_p \mathbf{m}_p,
\end{equation}
with the complex eigenvectors $\mathbf{m}_p$ and eigenvalues $\mu_p$ of $\mathrm{M}$. The collective polarizability
\begin{equation}
    \alpha_\mathrm{coll} = \sum_p \alpha_p = \sum_p \frac{1}{\mu_p}= \sum_p \frac{d_{ge}^2}{\hbar (\omega_0 - \omega + \Delta_p) - \mathrm{i}\hbar (\gamma_0 + \gamma_p)}
\end{equation}
is given by the polarizability $\alpha_p$ of each mode. The dipole-dipole interaction leads to a frequency shift $\Delta_p = d_{ge}^2 \mathrm{Re}(g_p)$ and broadening $\gamma_p = d_{ge}^2 \mathrm{Im}(g_p)$ of each mode, where $g_p$ are the complex eigenvalues of $\mathrm{G}$. The response of the dipole lattice to an external electric field is described by the collective dipole moment
\begin{equation}
    \mathbf{d} = \alpha_\mathrm{coll} \mathbf{E}_0 = \sum_p b_p \alpha_p \mathbf{m}_p,
\end{equation}
with expansion coefficients $b_p$ defined by a vector decomposition of the electric field $\mathbf{E}_0 = \sum_p b_p \mathbf{m}_p$. The collective response can be measured experimentally by the extinction of the dipole lattice
\begin{equation}
    \sigma_\mathrm{ext} = \frac{k}{\epsilon_0 E_0^2} \mathrm{Im}(\mathbf{E}_0^\dagger\cdot \mathbf{d}) \approx  \frac{k}{\epsilon_0 E_0^2} \sum_p \vert b_p \vert^2 \mathrm{Im}(\alpha_p).
\end{equation}

 For the simulations of the perfect lattice we assumed dipoles with $\hbar\omega_0=2\,$eV, $\hbar\gamma_0=5\,$meV, and $d_{ge}=10\,$D. The dipole polarization was along the $x$ axis (horizontal in the eigenvector plots of Fig.~\ref{fig:CollectiveDipoleTheory}). They were placed in an $N\times N$ square lattice with a lattice constant $a=1\,$nm. We calculated the eigenenergies, extinction spectra, and eigenvectors for $N=1-10$. The simulations for the experimental MePTCDI monolayer used $\hbar\omega_0=2.31\,$eV and $d_{ge}=8.8\,$D as measured in solution, see Supplementary Information. The broadening parameter was set to $\hbar\gamma_0=5\,$meV. We simulated a $20\times20$ lattice with $a=1.2\,$nm and the dipoles oriented at 45$^\circ$ to the lattice vectors which corresponds to the brick stone lattice. To account for the dielectric screening by the hBN and graphene substrates we averaged the dielectric contribution by vacuum on top ($\epsilon_m=1$) and by a half space filled with the van der Waals material underneath the MePTCDI layer. With a direction-averaged dielectric constant of $\epsilon_\mathrm{hBN}=4.3$ we obtain a background dielectric constant $\epsilon_m(\mathrm{hBN})=2.7$ for hBN and $\epsilon_\mathrm{G}=5.0$ leading to $\epsilon_m(\mathrm{G})=3.0$ for graphene.\cite{Laturia2018,Song2018} This approach is an upper bound for the screening by the environment. A smaller screening would increase the predicted frequency shift due to the formation of the collective state.

 For the lattices with inhomogeneously broadened dipole response we modeled the eigenstates and the extinction spectra for 2D lattices in which the individual dipole frequencies vary randomly $\omega_0 + \delta\omega_i$; $\delta\omega_i$ is obtained through random sampling from a Gaussian distribution with standard deviation $\sigma$. As for spatial disorder we report an average over many simulation runs. Average spectra are calculated from 5, 10, 15, and 20 random lattices where the frequencies of individual dipoles vary by $\sigma = 1\%$, $2\%$, $3\%$, and $4\%$ with respect to $\omega_0 = 2\,\mathrm{eV}$. 
Spatial disorder was simulated by varying the dipole positions $\mathbf{r}_i + \delta\mathbf{r}_i$ in the $xy$ plane around the sites $\mathbf{r}_i$ of the perfect lattice, where $\delta\mathbf{r}_i$ is obtained through random sampling from a multivariate Gaussian distribution and again averaged over several hundred runs. We performed similar simulations for variations in the dipole orientation (data not shown) with very similar results as for spatial disorder.
 
\subsection*{Exfoliation \& Physical Vapor Deposition} As substrates we used the freshly cleaved van der Waals materials hBN and graphite. They were exfoliated by the standard scotch tape approach onto a quartz wafer, resulting in few- and multi-layers with up to a few micrometer thickness. The out-baking process and the monolayer growth were performed in a tube furnace from Heraeus. After baking out the substrate with the 2D material at 400$^{\circ}$\,C for an hour to remove dirt and water residues on the substrate surface, the quartz wafer with the van der Waals material was placed at the end of the furnace where the furnace temperature starts to decrease. The N,N'-Dimethyl-3,4,9,10-Perylentetracarboxylicdiimide (MePTCDI) molecules from TCI were placed in an evaporation boat in the middle of the furnace. Argon flow (100\,sccm) and vacuum (50\,mbar) ensure that the molecules are transported to the growth substrate. The monolayer growth process took 90\,min. 

\subsection*{Fluorescence \& Raman Measurements} The fluorescence microscopy images are recorded with a Nikon Eclipse LV100 microscope. The microscope is equipped with a Nikon DS-Ri2 camera, a Thorlabs bandpass filter (FLH532-4) and a Thorlabs longpass filter (FELH0550) to filter out the desired wavelengths. \\
For the detection of the fluorescence and Raman spectra an XploRA (Horiba) Raman spectrometer was used. All fluorescence measurements were performed at 532\,nm laser excitation with 25\,$\mu W$ laser power and an integration time of 0.1\,s. The resonant Raman profile was recorded by tuning the excitation wavelength in steps of 10\,nm. Therefore we used an Ar-Kr ion-laser (Innova - Coherent) and a continuous ring laser from Radiant Dyes that can be operated with different fluorescent dyes (R6G and DCM) as lasing medium. A 100x objective (NA = 0.9) focused the laser beam on the sample. The Raman scattered light was detected in backscattering configuration by a Jobin-Yvon T64000 spectrometer in single mode (direct path) configuration. To detect the Raman scattered light the spectrometer is equipped with an Andor iDus CCD camera. All Raman spectra were fitted by a Lorenzian line shape. The intensity (integrated area under the peak) was plotted as a function of the excitation wavelength. To account for wavelength dependent changes in the sensitivity of the Raman setup the Raman intensity was calibrated on benzonitrile that has a constant Raman cross section.\cite{Piao2016,Duque2011} All fluorescence and Raman measurements were performed under ambient conditions.

\subsection*{Micro-Absorbance}
The absorption spectra are recorded under ambient conditions with a home built micro-absorbance setup \cite{Mueller2018, Mueller2020} that is equipped with a broadband light source (NKT - FIU 15). The light is guided to an inverse microscope from Olympus where the light is focused by an 100x objective (NA = 0.9) onto the sample. The transmitted light (T) is collected by a second 100x objective (NA = 0.8) and afterwards guided through an optical fiber to an Avantes spectrometer. The reflected light (R) goes through a beam splitter, is collected by a collimator lens that couples the light into a fiber that guides the light to the spectrometer. The absorbance (A) is calculated from the reflection and transmission spectra as A~=~100~\%~-~R~-~T.

\subsection*{AFM Measurements} The AFM images were taken with an Cypher ES atomic force microscope from Asylum Research Oxford Instruments Inc. To remove possible organic contaminations of the tip, the AFM cantilevers were cleaned with argon plasma with Zepto instrument (Diener electronics Inc.) at 50\% power for one minute. AFM imaging was performed in amplitude modulation, called also tapping mode. The cantilever (qp-fast, 15\,N/m, Nanosensors) was excited in resonance with its third eigenmode around 4.5\,MHz with an amplitude of less than one nanometer. Further details of atomic resolution with high-eigenmode tapping mode atomic force microscopy are described in Ref.~\cite{Severin2022}. The AFM cell was at room temperature and was continuously purged with dry nitrogen.

\begin{acknowledgement}

We thank S. Kirstein for a critical reading of this manuscript and useful discussion. This work was supported by the European Research Council (ERC) under grant DarkSERS-772\,108, the German Science Foundation (DFG) under grant Re2654/13 and Re2644/10, and the SupraFAB Research Center at Freie Universit\"at Berlin. N.S.M. acknowledges support from the German National Academy of Sciences Leopoldina through the Leopoldina Postdoc Scholarship. 

\end{acknowledgement}

\newpage
\renewcommand{\figurename}{Supplementary Fig.}
\setcounter{figure}{0}

\section{Supporting Information}

\subsection{Simulations of Dipole Chains (1D)}
\begin{figure}
\centering
\includegraphics[width=0.67\textwidth]{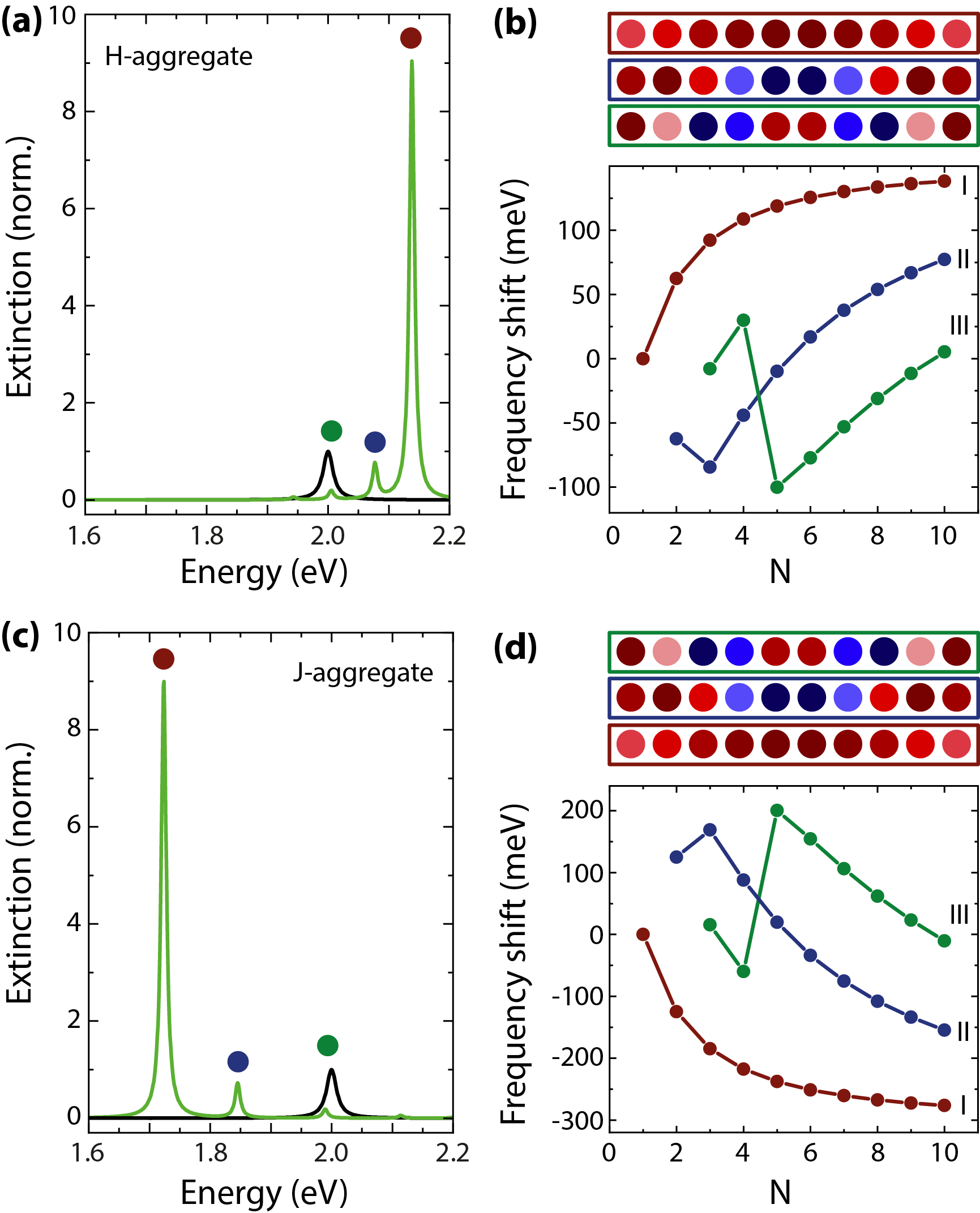}
\caption{\label{fig:Chain} Collective extinction of a 1D dipole chain. \textbf{(a,c)} collective extinction of a 1D dipole chain with dipole orientation in $y$-direction (H-aggregate) and along $x$ (J-aggregate). Showing the typical blue/red shift. The colored dotes indicate to which eigenmode in (b,d) the spectrum belongs to. \textbf{(b,d)} Energy shift of selected eigenstates with eigenvectors shown above. Eigenstate I (red curve) has the strongest dipole moment and is optical active. For the eigenvectors, red and blue correspond to opposite orientations of transition dipoles.}
\end{figure}
\newpage

\subsection{MePTCDI monolayer Height Determination by AFM}
\begin{figure}
\centering
\includegraphics[width=0.7\textwidth]{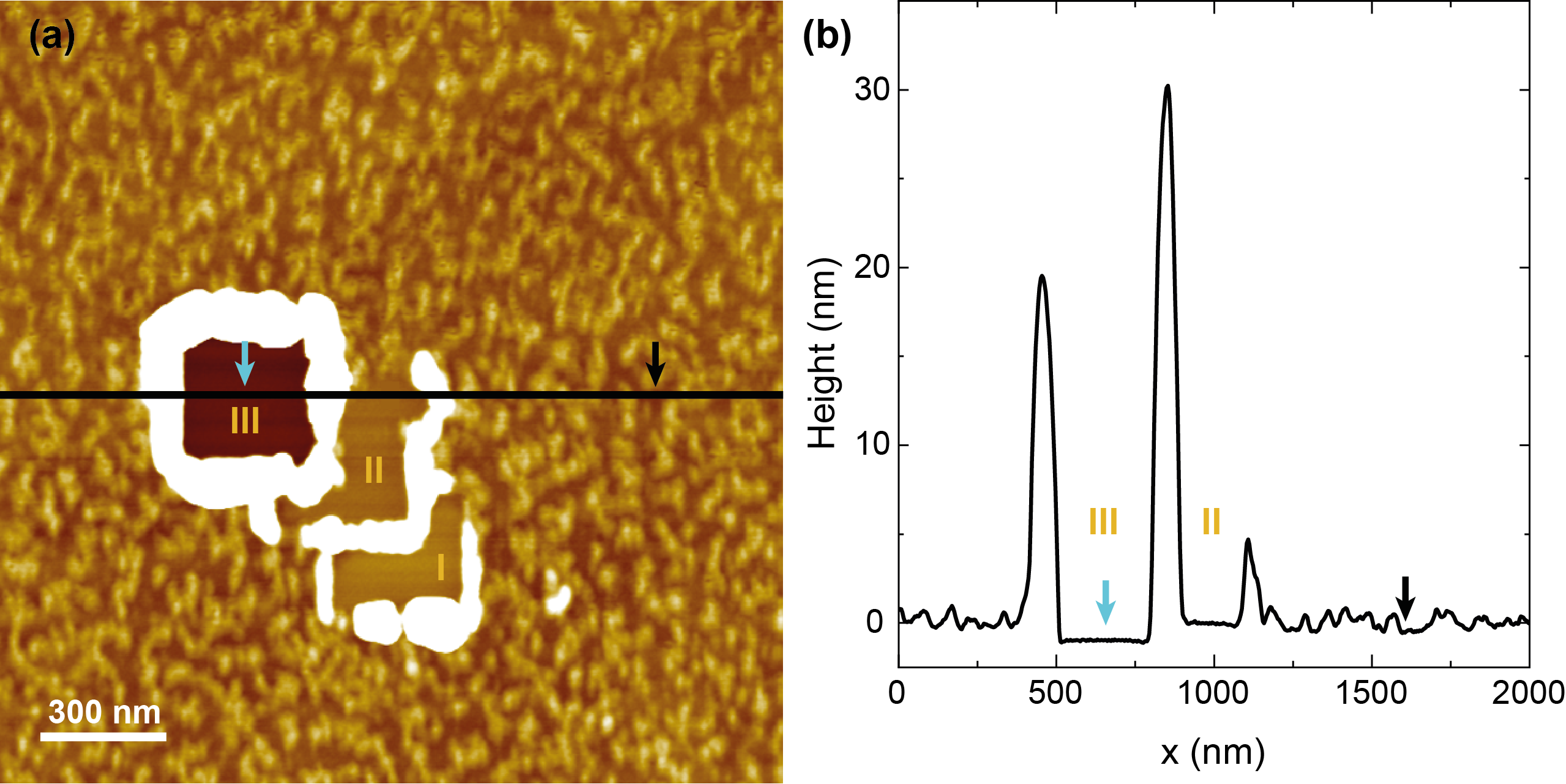}
\caption{\label{fig:AFM_SI} Determination of the organic monolayer height on hBN. \textbf{(a)} Large scale AFM image showing three areas (I-III) which were imaged in contact mode with progressively larger force from III to I. Area III is lower than the surrounding areas by about 0.5\,nm. \textbf{(b)} Height profile along the black line in panel (a). The height difference
between area I (blue arrow) and the area between the agglomerates (black arrow) is approximately 0.5\,nm which is in good agreement with the height of a MePTCDI monolayer.}
\end{figure}

In order to try to assess the thickness of the molecular layer, we attempted to clean a small area with an AFM tip. For this the imaging was switched to contact mode, and a $300~\times~300$\,nm$^2$ area was repeatedly imaged. Then the imaging was switched to tapping mode and an $2\times2\,\mu m^2$ large overview image was taken to assess the topography changes produced by the contact mode imaging, Supplementary Fig.~\ref{fig:AFM_SI}a. We used 160AC-NG cantilever (Opus by Mikromasch) with spring constant 26\,N/m as specified by the manufacturer. While imaging in contact mode with intermediate forces allowed to rub away molecular piles, apparently it did not allow to clean up the molecular layer. Just imaging in contact mode with a rather large force (about 11\,$\mu$N) resulted in reduced height in the imaged area. The height reduction matches roughly the expected height of a molecular monolayer, Supplementary Fig.~\ref{fig:AFM_SI}b. The force was calculated assuming the manufacturer specified spring constant, the deflection sensitivity was calculated from a force distance curve acquired on the same sample. Typically forces required to rub away molecular layers are smaller.\cite{Yagodkin2022} We note that even if the tip removed the top hBN layer, it is not expected that the freshly exposed hBN surface is covered with the molecules. Therefore, the result supports our conclusion on the monolayer thickness of the molecular layer. 

\newpage

\subsection{Emission Spectrum of MePTCDI Agglomerates on hBN}
\begin{figure}[h]
\centering
\includegraphics[width=0.26\textwidth]{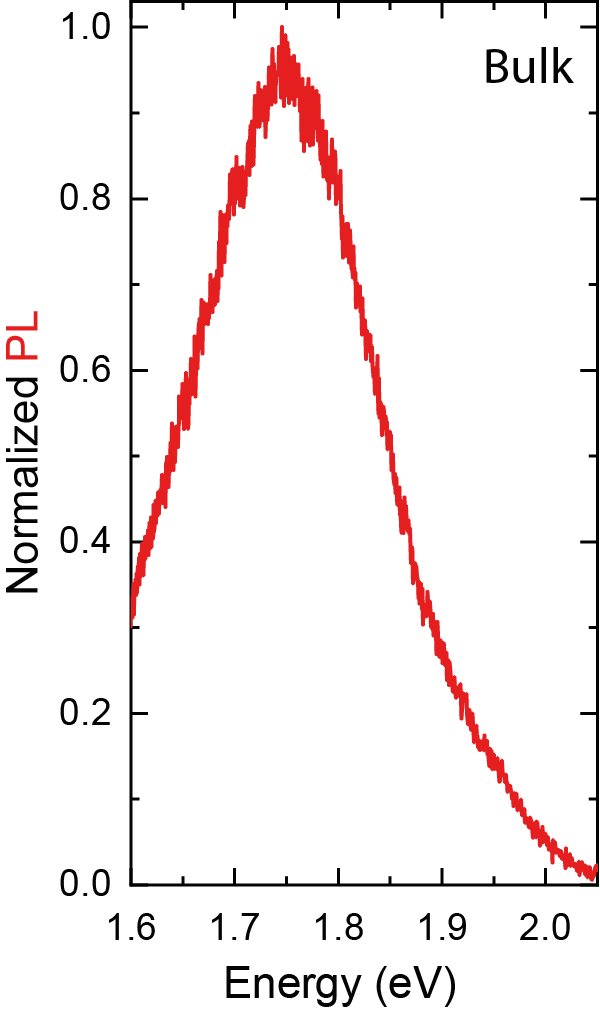}
\caption{\label{fig:Bulk} Micro-fluorescence spectrum of MePTCDI agglomerates. Broad fluorescence spectrum of a bulk/crystalline area of MePTCDI molecules in Fig.~3a. The spectrum was measured by exciting the structure with a 532\,nm laser.}
\end{figure}

\newpage

\subsection{Large Scale AFM Images of MePTCDI on hBN} 
\begin{figure}[h]
\centering
\includegraphics[width=0.78\textwidth]{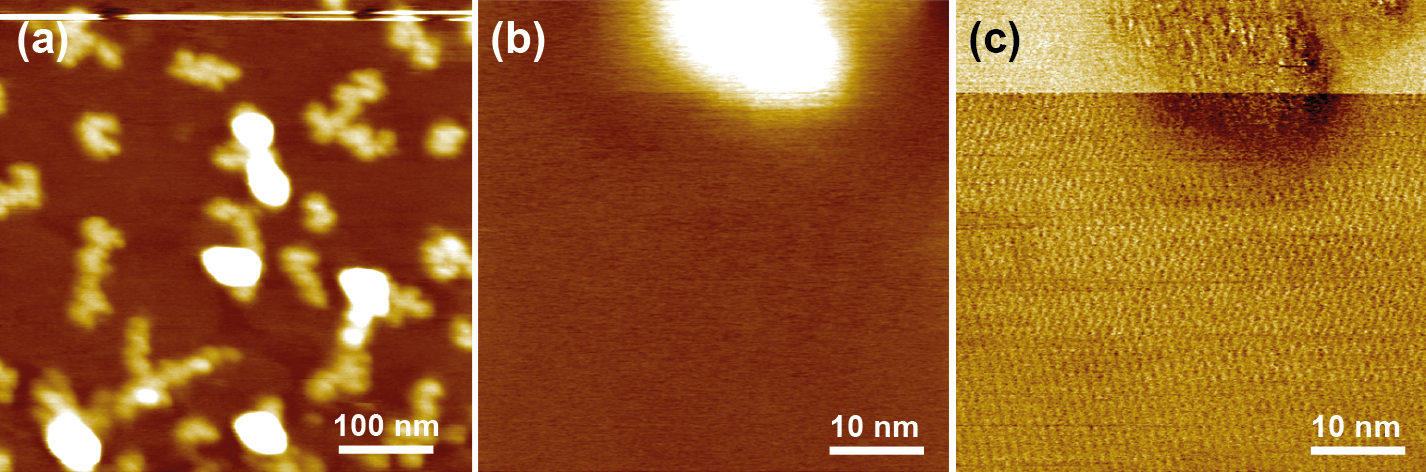}
\caption{\label{fig:AFMOver} Overview AFM images of the MePTCDI on hBN. \textbf{(a)} AFM topography image of a sample area where some agglomerates of 2~-~3\,nm height start to grow. \textbf{(b)}~Zoom in a flat area of panel (a). The topography shows a homogeneous sample area. \textbf{(c)}~Phase image taken simultaneously with the topography image in (b), giving the reported molecular brick stone lattice in Fig.~3 of the main manuscript. The images were taken at the beginning of an imaging session. Therefore, the images were slightly distorted by piezo drift and they were not included into the unit cell statistics.}
\end{figure}

\newpage

\subsection{Transition Dipole Moment of MePTCDI}
\label{TransitionDipoleMoment}

To calculate the transition dipole moment ($\mu$)
\begin{equation}
    \mu = \sqrt{\frac{3\hbar e^2}{2m_e\omega_0}f} = 8.8\,D,
\end{equation}
the oscillator strength (f) was estimated by
\begin{equation}
    f = \frac{4m_ec\epsilon_0}{N_Ae^2}ln(10)\int\epsilon(\nu)d\nu = 0.68,
\end{equation}
where $\hbar$ is the reduced Planck constant, $e$ the electric charge, $m_e$ the electron mass, $\omega_0$ the transition frequency, $c$ the speed of light, $\epsilon_0$ is the electrostatic constant, $N_A$ the Avogadro constant, and $\epsilon(\nu)$ molar absorption coefficient. The $\epsilon(\nu)$ in Supplementary Fig.~\ref{fig:Screening} was determined by the absorption of the MePTCDI molecules dissolves in chloroform.

\begin{figure}[h]
\centering
\includegraphics[width=0.4\textwidth]{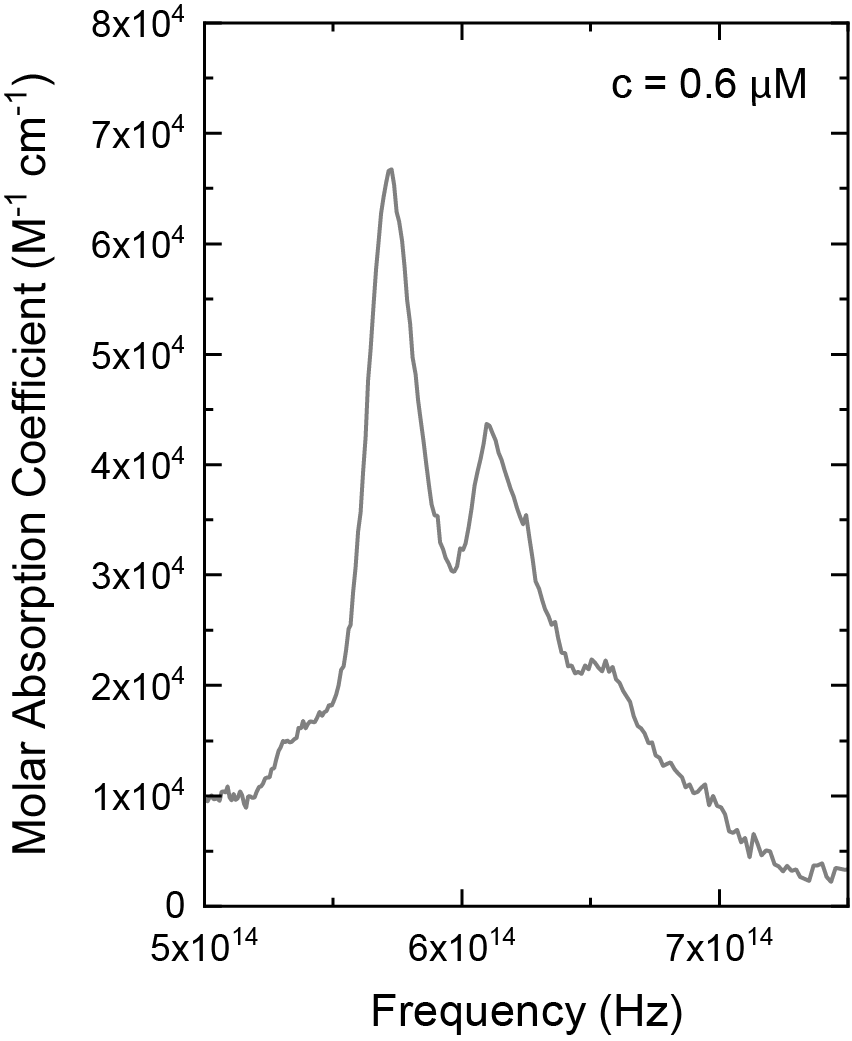}
\caption{\label{fig:Screening} Molar absorption coefficient of MePTCDI dissolved in chloroform. The molar absorption coefficient $\epsilon(\nu)$ was calculated from the absorption spectrum. The concentration of the solution was 0.6\,$\mu M$ and the path length 1\,cm.}
\end{figure}

\bibliography{MePTCDI-bib_final_ganz}

\end{document}